\documentclass[aps,preprint]{revtex4}%
\usepackage{amsfonts}
\usepackage{amsmath}
\usepackage{amssymb}
\usepackage{graphicx}%
\begin{document}
\title[Vortices in the BEC-BCS crossover regime]{The vortex state in the BEC to BCS crossover: a path-integral description }
\author{J. Tempere}
\affiliation{TFVS, Departement Fysica, Universiteit Antwerpen, Universiteitsplein 1, B2610
Antwerpen, Belgium.}
\author{M. Wouters}
\affiliation{TFVS, Departement Fysica, Universiteit Antwerpen, Universiteitsplein 1, B2610
Antwerpen, Belgium.}
\author{J. T. Devreese}
\affiliation{TFVS, Departement Fysica, Universiteit Antwerpen, Universiteitsplein 1, B2610
Antwerpen, Belgium.}

\begin{abstract}
We derive a path-integral description of the vortex state of a fermionic
superfluid in the crossover region between the molecular condensate (BEC)
regime and the Cooper pairing (BCS) regime. This path-integral formalism,
supplemented by a suitable choice for the saddle point value of the pairing
field in the presence of a vortex, offers a unified description that
encompasses both the BEC and BCS limits. The vortex core size is studied as a
function of the tunable interaction strength between the fermionic atoms. We
find that in the BEC regime, the core size is determined by the molecular
healing length, whereas in the BCS regime, the core size is proportional only
to the Fermi wave length. The observation of such quantized vortices in dilute
Fermi gases would provide an unambiguous proof of the realization of
superfluidity in these gases.

\end{abstract}
\date{10/10/2004}
\maketitle

\section{Introduction}

The realization of superfluidity in dilute fermionic gases
\cite{GreinerNAT426} has opened up a new avenue for the investigation of
fermionic quantum systems. In such gases, the presence of Feshbach resonances
in the interatomic scattering allows to tune the interaction strength from
strongly repulsive to strongly attractive \cite{BartensteinPRL92}. In the
limit of weak repulsive interactions, deeply bound molecular bosons are formed
that Bose condense \cite{GreinerNAT426} (BEC). In the limit of weak attractive
interactions, a Bardeen-Cooper-Schrieffer (BCS) fermionic superfluid arises
\cite{RegalPRL92}. The crossover between both regimes has recently attracted a
great deal of theoretical and experimental interest.

One of the hallmarks of superfluidity, be it bosonic or fermionic, is the
presence of quantized vortices. Whereas vortices are well understood both in
the BEC and the BCS limits, it is not clear how the characteristics of the
vortex (such as the core size) behave in the crossover regime. Bulgac and Yu
have extended density functional theory to superfluid fermion systems
\cite{BulgacPRL88}, and studied vortex states in the BCS regime within their
superfluid local density approximation (SLDA) \cite{BulgacPRL91}. They found
that in the BCS regime vortices give rise to a depletion in the fermion
density. These results confirm earlier calculations based on the Bogoliubov-de
Gennes theory \cite{NygaardPRL90}.

The goal of this communication is to develop a path-integral treatment suited
to describe vortices in the BCS, BEC, and crossover regimes. With this
treatment, we investigate how the vortex core size and the fermionic density
depletion at the core change when the fermionic superfluid is brought from the
BEC to the BCS regime. On the BCS side of the Feshbach resonance, we compare
our results to those obtained with the SLDA treatment \cite{BulgacPRL91}.

A path-integral treatment for the ground state of the fermionic superfluid was
developed by S\'{a} de Melo, Randeria and Engelbrecht
\cite{RanderiaPRL71,RanderiaPRB55}. Their formalism provides a unified
description of the BEC, BCS, and crossover regimes and predicts a smooth
cross-over for the critical temperature, pairing gap, and chemical potential.
They consider a homogeneous Fermi gas of atoms determined by the action
functional
\begin{align}
\mathcal{S}_{1}  &  =\int\limits_{0}^{\beta}d\tau\int d\mathbf{x}\text{ }%
{\textstyle\sum_{\sigma}}
\left[  \bar{\psi}_{\mathbf{x},\tau,\sigma}\left(  \frac{\partial}%
{\partial\tau}-\frac{1}{2m}\mathbf{\nabla}_{\mathbf{x}}^{2}-\mu\right)
\psi_{\mathbf{x},\tau,\sigma}\right] \nonumber\\
&  +\int\limits_{0}^{\beta}d\tau\int d\mathbf{x}\text{ }g\bar{\psi
}_{\mathbf{x},\tau,\uparrow}\bar{\psi}_{\mathbf{x},\tau,\downarrow}%
\psi_{\mathbf{x},\tau,\downarrow}\psi_{\mathbf{x},\tau,\uparrow}. \label{S1}%
\end{align}
In this expression $\bar{\psi}_{\mathbf{x},\tau,\sigma}$ and $\psi
_{\mathbf{x},\tau,\sigma}$ are the Grassmann variables describing the
fermionic degrees of freedom, where $\mathbf{x}$ is the position vector,
$\tau$ the imaginary time and $\sigma=\uparrow,\downarrow$ denotes the two
hyperfine spin states present in the Fermi gas. The chemical potential is
denoted by $\mu$, and $\beta=1/(k_{B}T)$ is the inverse temperature$.$ The
interaction between the fermionic atoms only takes place between atoms in
different hyperfine spin states, and is described by a contact interaction
characterized by the renormalized strength $g$. The partition sum is given by
the functional integral over the Grassmann variables,
\begin{equation}
\mathcal{Z}=\int\mathcal{D}[\bar{\psi}_{\mathbf{x},\tau,\sigma},\psi
_{\mathbf{x},\tau,\sigma}]\exp\left\{  -\mathcal{S}_{1}\right\}  .
\end{equation}
To unravel the product of four Grassmann variables, the Hubbard-Stratonovic
transformation is performed. This transformation introduces the bosonic
Hubbard-Stratonovic fields $\Delta_{\mathbf{x},\tau}$ and $\bar{\Delta
}_{\mathbf{x},\tau}$ such that%
\begin{equation}
\mathcal{Z}=\int\mathcal{D}[\bar{\Delta}_{\mathbf{x},\tau},\Delta
_{\mathbf{x},\tau}]\int\mathcal{D}[\bar{\psi}_{\mathbf{x},\tau,\sigma}%
,\psi_{\mathbf{x},\tau,\sigma}]\exp\left\{  -\mathcal{S}_{2}\right\}  .
\end{equation}
with%
\begin{align}
\mathcal{S}_{2}  &  =%
{\displaystyle\int\limits_{0}^{\beta}}
d\tau\int d\mathbf{x}\text{ }\left\{
{\textstyle\sum_{\sigma}}
\hbar\beta\left[  \bar{\psi}_{\mathbf{x},\tau,\sigma}\left(  \frac{\partial
}{\partial\tau}-\frac{1}{2m}\mathbf{\nabla}_{\mathbf{x}}^{2}-\mu\right)
\psi_{\mathbf{x},\tau,\sigma}\right]  \right. \nonumber\\
&  \left.  -\bar{\Delta}_{\mathbf{x},\tau}\psi_{\mathbf{x},\tau,\downarrow
}\psi_{\mathbf{x},\tau,\uparrow}-\Delta_{\mathbf{x},\tau}\bar{\psi
}_{\mathbf{x},\tau,\uparrow}\bar{\psi}_{\mathbf{x},\tau,\downarrow}-\frac
{\bar{\Delta}_{\mathbf{x},\tau}\Delta_{\mathbf{x},\tau}}{g}\right\}  .
\label{S2}%
\end{align}
The action functional $\mathcal{S}_{2}$ is quadratic in the Grassmann
variables so that the functional integration over these variables can in
principle be evaluated.

\section{Saddle point for the vortex state}

In practice, to perform the functional integration over Grassmann variables,
one has to choose a saddle point value for the fields $\bar{\Delta
}_{\mathbf{x},\tau},\Delta_{\mathbf{x},\tau}$ . To describe the \emph{ground
state}, S\'{a} de Melo et al. \cite{RanderiaPRL71}\ suitably choose a uniform
constant saddle point $\left(  \bar{\Delta}_{\mathbf{x},\tau}\right)  ^{\ast
}=\Delta_{\mathbf{x},\tau}=\left\vert \Delta\right\vert $. After doing this,
the Grassmann variables can be integrated out straightforwardly, resulting in
an effective saddle-point action%
\begin{equation}
\mathcal{S}_{\text{eff}}=%
{\displaystyle\int\limits_{0}^{\beta}}
d\tau\int d\mathbf{x}\text{ }\left\{  -\frac{1}{\beta}\operatorname*{tr}%
\left[  \ln\left(  -\mathbb{G}^{-1}\right)  \right]  -\frac{\left\vert
\Delta\right\vert ^{2}}{g}\right\}  , \label{Seff}%
\end{equation}
where $\mathbb{G}^{-1}$ is the inverse Nambu propagator, for the ground state:%
\begin{equation}
-\mathbb{G}_{\text{ground state}}^{-1}=\sigma_{0}\frac{\partial}{\partial\tau
}-\sigma_{1}\left\vert \Delta\right\vert -\sigma_{3}\left[  \frac{1}%
{2m}\mathbf{\nabla}_{\mathbf{x}}^{2}+\mu\right]  ,
\end{equation}
and the $\sigma_{j}$ are Pauli matrices. The saddle point equation
$\delta\mathcal{S}_{\text{eff}}/\delta\left\vert \Delta\right\vert =0$ then
leads to the familiar gap equation for the fermionic superfluid
\cite{HeiselbergPRL85}. The chemical potential $\mu$ is fixed by the fermion
density. At finite temperatures, fluctuations around the saddle point value
can be taken into account to improve the theory and find the critical
temperature \cite{RanderiaPRL71,RanderiaPRB55}.

To investigate the \emph{vortex state}, we propose to use a different saddle
point, and set
\begin{equation}
\left(  \bar{\Delta}_{\mathbf{x},\tau}\right)  ^{\ast}=\Delta_{\mathbf{x}%
,\tau}=\left\vert \Delta_{r}\right\vert \exp(i\theta), \label{vortsadlpt}%
\end{equation}
where $\theta$ is the angle around the vortex line and $r$ is the distance to
the vortex line. That is, $r$ and $\theta$ are the radial and angular
coordinates if one chooses cylindrical coordinates $\mathbf{x}=(r,\theta,z)$
such that the $z$-axis lies along the (straight) vortex line. This particular
choice of the saddle point value lies at the core of the present treatment.
With this choice for the saddle point value, the integration over Grassmann
variables leads again to an effective action of the form (\ref{Seff}), but
with a different result for $\mathbb{G}^{-1}$. We find:%
\begin{align}
-\mathbb{G}_{\text{vortex}}^{-1}  &  =\sigma_{0}\left[  \frac{\partial
}{\partial\tau}-\frac{i}{2mr}\mathbf{e}_{\phi}\cdot\mathbf{\nabla}%
_{\mathbf{x}}\right]  -\sigma_{1}\left\vert \Delta_{r}\right\vert \nonumber\\
&  +\sigma_{3}\left[  -\frac{1}{8mr^{2}}-\frac{1}{2m}\mathbf{\nabla
}_{\mathbf{x}}^{2}-\mu\right]  . \label{Gvortex}%
\end{align}
The density of paired atoms near the vortex core ($\left\vert \Delta
_{r}\right\vert $ as a function of $r$) can in principle be derived from the
saddle-point equation $\delta\mathcal{S}_{\text{eff}}/\delta\left\vert
\Delta_{r}\right\vert =0$ (although in practice the presence of the spatial
derivatives inhibits straightforward calculation). However, the result no
longer contains explicit information on the density of fermionic atoms. The
results of Bulgac \emph{et al.} interestingly show that whereas $\left\vert
\Delta_{r}\right\vert $ tends to zero at the vortex core, the fermionic
density needs not go to zero. To study the density of fermionic atoms,
expression (\ref{Gvortex}) is not suited.

\section{Expliciting the fermionic density}

In order to introduce the fermionic atom density in the path-integral
expressions, a transformation was proposed by De Palo et al.
\cite{DePaloPRB60} based on the identity
\begin{align}
C  &  =\int\mathcal{D}\left[  \rho_{\mathbf{x},\tau}^{HS},\rho_{\mathbf{x}%
,\tau}\right]  \exp\left[  -%
{\displaystyle\int\limits_{0}^{\beta}}
d\tau\int d\mathbf{x}\;i\rho_{\mathbf{x},\tau}^{HS}\right. \\
&  \left.  \times\left(  \rho_{\mathbf{x},\tau}-\bar{\psi}_{\mathbf{x}%
,\tau,\uparrow}\psi_{\mathbf{x},\tau,\uparrow}-\bar{\psi}_{\mathbf{x}%
,\tau,\downarrow}\psi_{\mathbf{x},\tau,\downarrow}\right)  \right]
\label{Cdepalo}%
\end{align}
where $C$ is a constant c-number and $\rho_{\mathbf{x},\tau}^{HS}%
,\rho_{\mathbf{x},\tau}$ are bosonic fields. This generalized delta-function
expression identifies $\rho_{\mathbf{x},\tau}$ with $%
{\textstyle\sum_{\sigma}}
\bar{\psi}_{\mathbf{x},\tau,\sigma}\psi_{\mathbf{x},\tau,\sigma}$, the
fermionic density.

Multiplying the partition sum (\ref{S2}) with the constant (\ref{Cdepalo}),
and approximating the Hubbard-Stratonovich fields $\bar{\Delta}_{\mathbf{x}%
,\tau},\Delta_{\mathbf{x},\tau}$ by the vortex saddle point (\ref{vortsadlpt})
leads to the following saddle-point approximation for the partition sum%
\begin{equation}
\mathcal{Z}_{\text{sp}}=\int\mathcal{D}\left[  \bar{\psi}_{\mathbf{x}%
,\tau,\sigma},\psi_{\mathbf{x},\tau,\sigma}\right]  \int\mathcal{D}\left[
\rho_{\mathbf{x},\tau}^{HS},\rho_{\mathbf{x},\tau}\right]  \text{ }%
\exp\left\{  -\mathcal{S}_{\text{sp}}\right\}  , \label{Zsp}%
\end{equation}
with
\begin{align}
\mathcal{S}_{\text{sp}}  &  =%
{\displaystyle\int\limits_{0}^{\beta}}
d\tau\int d\mathbf{x}\text{ }\left\{
{\textstyle\sum_{\sigma}}
\left[  \bar{\psi}_{\mathbf{x},\tau,\sigma}\left(  \frac{\partial}%
{\partial\tau}+\frac{1}{8mr^{2}}+\right.  \right.  \right. \nonumber\\
&  \left.  \left.  -\frac{1}{2m}\mathbf{\nabla}_{\mathbf{x}}^{2}-\mu
-i\rho_{\mathbf{x},\tau}^{HS}\right)  \psi_{\mathbf{x},\tau,\sigma}\right]
+i\rho_{\mathbf{x},\tau}^{HS}\rho_{\mathbf{x},\tau}\nonumber\\
&  \left.  -\left\vert \Delta_{r}\right\vert (\psi_{\mathbf{x},\tau
,\downarrow}\psi_{\mathbf{x},\tau,\uparrow}+\bar{\psi}_{\mathbf{x}%
,\tau,\uparrow}\bar{\psi}_{\mathbf{x},\tau,\downarrow})-\frac{\left\vert
\Delta_{r}\right\vert ^{2}}{g}\right\}  . \label{Ssp}%
\end{align}
To get rid of the path integration over $\rho_{\mathbf{x},\tau}^{HS}%
,\rho_{\mathbf{x},\tau}$ in (\ref{Zsp}), we again use a saddle-point approach
and set these fields equal to $\rho_{\mathbf{x},\tau}^{HS}=\rho_{r}^{HS},$
$\rho_{\mathbf{x},\tau}=\rho_{r}$. This implies that the fermion density only
depends on the distance $r$ from the vortex core (and not on the $\theta$ or
$z$ coordinate). From (\ref{Ssp}) it is clear that the auxiliary field
$i\rho_{r}^{HS}$ is related to the chemical potential. It is useful to
introduce \cite{DePaloPRB60}%

\begin{equation}
\zeta_{r}=i\rho_{r}^{HS}+\mu-\frac{1}{8mr^{2}}.
\end{equation}
On the level of the fermionic degrees of freedom, $\zeta_{r}$ acts as a
one-body potential combining the auxiliary field, the chemical potential and
the angular momentum barrier. Alternatively, it can be interpreted as a local
chemical potential.

To perform the integration over the Grassmann variables we assume that
$\zeta_{r}$ and the pairing field $\Delta_{r}$ vary slowly in comparison with
the relevant fermion frequencies. After integration over $\bar{\psi
}_{\mathbf{x},\tau,\sigma},\psi_{\mathbf{x},\tau,\sigma}$, we then obtain the
result $\mathcal{Z}_{\text{sp}}\propto\exp\left\{  \mathcal{S}_{\text{eff}%
}^{\prime}\right\}  $ with
\begin{align}
\mathcal{S}_{\text{eff}}^{\prime}  &  =-\left\{  \frac{\left\vert \Delta
_{r}\right\vert ^{2}}{g}-\left(  \zeta_{r}-\mu+\frac{\hbar^{2}}{8mr^{2}%
}\right)  \rho_{r}\text{ }\right. \nonumber\\
&  \left.  +2\int\tfrac{d\mathbf{k}}{(2\pi)^{3}}\ln\left[  2\cosh\left(
\tfrac{\beta}{2}\sqrt{\left(  \tfrac{k^{2}}{2m}-\zeta_{r}\right)
^{2}+\left\vert \Delta_{r}\right\vert ^{2}}\right)  \right]  \right\}  .
\end{align}

\section{Properties of the vortex core}

The first saddle point equation is
\begin{equation}
\frac{\partial\mathcal{S}_{\text{eff}}^{\prime}}{\partial\rho_{r}%
}=0\Leftrightarrow\zeta_{r}=\mu-\frac{1}{8mr^{2}}. \label{SPE1}%
\end{equation}
This locks the local chemical potential $\zeta_{r}$ as a function of the
overall chemical potential and the energy barrier of the vortex flow, as one
would expect in the Thomas-Fermi approximation in the molecular limit. The
parameter $\mu$ is determined by fixing the total number of particles (or the
density far away from the vortex core). Differentiating $\mathcal{S}%
_{\text{eff}}^{\prime}$ with respect to $\left\vert \Delta_{r}\right\vert $ we
find the familiar gap equation, but with a local chemical potential determined
by $\zeta_{r}$:
\begin{align}
\frac{1}{k_{F}a_{s}}  &  =-\frac{2}{\pi}\int dk\text{ }k^{2}\left(
\frac{\tanh(\beta E_{k}/2)}{E_{k}}-\frac{1}{k^{2}}\right)  ,\label{SPE2}\\
\text{with }E_{k}  &  =\sqrt{\left(  k^{2}-\zeta_{r}\right)  ^{2}+\Delta
_{r}^{2}}.\nonumber
\end{align}
Finally, the third saddle point equation $\partial\mathcal{S}_{\text{eff}%
}^{\prime}/\partial\zeta_{r}=0$ allows to calculate the fermionic density near
the vortex core%
\begin{equation}
\rho_{r}=\frac{3}{2}\int dk\text{ }k^{2}\left[  1-\frac{k^{2}-\zeta_{r}}%
{E_{k}}\tanh(\beta E_{k}/2)\right]  . \label{SPE3}%
\end{equation}
In (\ref{SPE2}),(\ref{SPE3}), units are such that wave numbers are expressed
in Fermi wave numbers and energies in Fermi energies.%

\begin{figure}
[ptb]
\begin{center}
\includegraphics[
height=3.0224in,
width=4.3286in
]%
{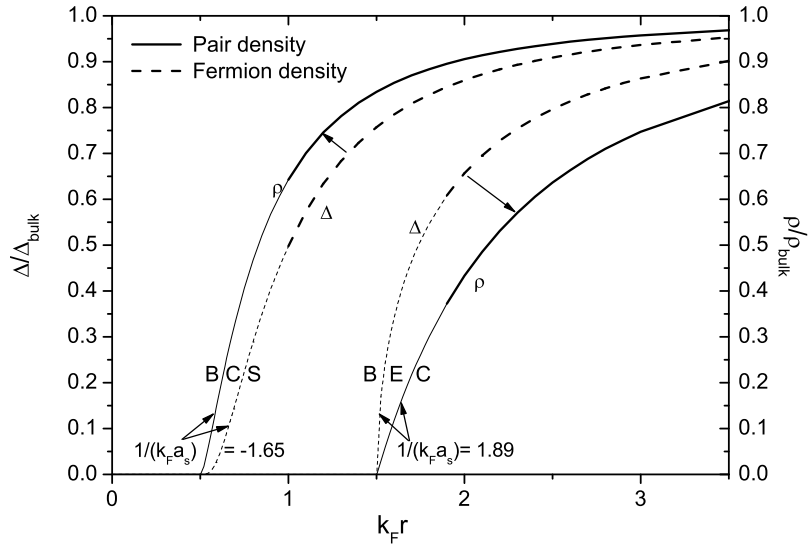}%
\caption{The pair density relative to the bulk pair density (left axis, dashed
curves) and the fermionic atom density relative to its bulk value (right axis,
full curves) are shown as a function of the distance to the vortex line, for
two different values of the interaction strength. The depletion of the fermion
density at the core is overestimated by the current approach, as discussed in
the text. The thin part of the curves indicate the region where the current
approach is unreliable.}%
\end{center}
\end{figure}

We investigate the solutions of these equations in the zero temperature limit
($\beta\rightarrow\infty$), so that we may assume that fluctuation corrections
around the saddle point values are not important. The result for the pair
density and the fermion density are shown in figure 1. Both the pair density
and the fermion density go to zero in a region with spacial dimensions of the
order of $1/k_{F}$ around the vortex line ($r=0$). In the BEC limit, the pair
density and the fermionic atom density become zero at the same distance from
the vortex line, so that no atoms are seen inside the core. This corresponds
to the expectation for a molecular BEC. In the BCS limit, fermionic atoms can
penetrate into the region where the pair density is zero. These results
qualitatively agree with those of Bulgac et al. \cite{BulgacPRL91} in the BCS
region, but quantitatively they are quite different: in refs.
\cite{BulgacPRL91,NygaardPRL90} only a small depression in the fermionic
density is found on the BCS side of the Feshbach resonance. What can be the
reason for the discrepancy ? Most likely the assumption that the pairing field
$\Delta_{r}$ is smooth on the length scale of the relevant fermion frequencies
is too crude. As in the case of the Thomas-Fermi approximation, which breaks
down in the region where the order parameter becomes zero, we expect that the
assumption that $\Delta_{r}$ varies smoothly breaks down in a region near the
point where $\Delta_{r}$ becomes zero. In figure 1, the results deemed
unreliable are shown in thin curves, whereas the results in the region where
the aforementioned assumption is estimated to hold\ are shown in thick curves.

Thus, we must conclude that this assumption may bring us qualitative insight
into the core region, but that the quantitative analysis of how exactly
$\Delta_{r}$ and $\rho_{r}$ depend on $r$ near the vortex core is misleading.
It is not yet clear whether the current result obtained by the saddle-point
approximation can be retrieved in the framework of a Bogoliubov-de Gennes
theory where similar assumptions (i.e. that $\Delta_{r}$ varies smoothly on
the length scale corresponding to the relevant fermion frequencies) are made.%

\begin{figure}
[ptb]
\begin{center}
\includegraphics[
natheight=2.995000in,
natwidth=3.852600in,
height=3.0653in,
width=3.935in
]%
{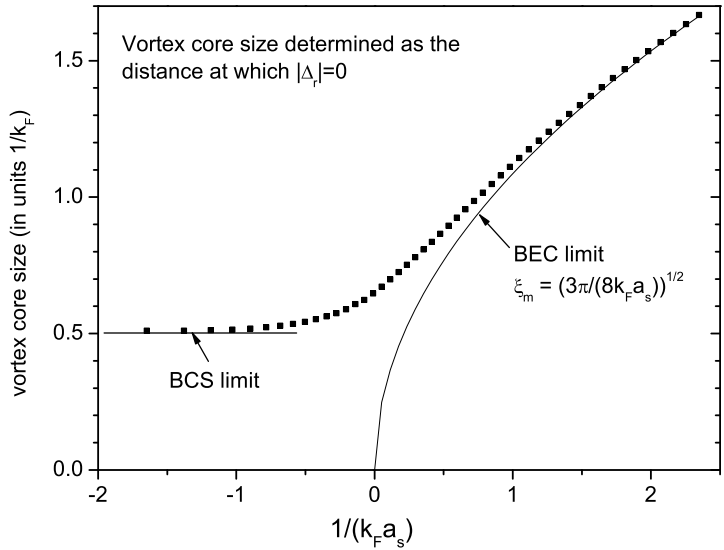}%
\caption{The vortex core size, in units $1/k_{F}$, is shown as a function of
the interaction strength $1/(k_{F}a_{s})$. In the BEC\ regime, the curve shows
the molecular healing length, whereas in the BCS limit the result goes to a
constant as indicated by the line.}%
\end{center}
\end{figure}

Still, the present results allow us to extract a value for the core size,
since for this purpose one does not need to know the exact functional
dependence of $\Delta_{r}$ on $r$. We estimate the size of the vortex core as
the distance from the core at which our solution for $\Delta_{r}$ becomes
zero. The result is shown in figure 2. In the molecular BEC limit, one expects
that the core size is related to the healing length for the molecular
Bose-Einstein condensate, $\xi_{m}=1/\sqrt{8\pi n_{m}a_{m}}$. In the
path-integral treatment to lowest order in the fluctuations around the saddle
point \cite{RanderiaPRB55}, the molecular scattering length is twice the
atomic scattering length, $a_{m}=2a_{s}$. The density of molecules is half the
density of atoms if all atoms form molecules, $n_{m}=k_{F}^{3}/(6\pi^{2})$.
Thus $\xi_{m}=\sqrt{3\pi/(8k_{F}a_{s})}$ if we express $\xi_{m}$ in units of
$k_{F}^{-1}$. Surprisingly, in the BEC\ limit we find that the vortex core
size derived with the path-integral method follows exactly the molecular
length $\xi_{m}$. Approaching the crossover region, the vortex core size
deviates from its molecular value. In the BCS limit, the vortex core size
tends to a constant, given by $1/(2k_{F})$, as indicated in figure 2. This
suggests that for vortices in ultracold dilute Fermi gases the vortex cores in
the BCS limit can be much smaller than the BCS correlation length
$\xi_{\text{BCS}}$ (since $k_{F}\xi_{\text{BCS}}\sim E_{F}/\Delta\gg1$), as is
the case superfluid vortices in neutron stars \cite{ElgaroyAA370}.

\section{Conclusions}

The path-integral formalism allows to investigate the ground state of the
fermionic superfluid both in the BCS, BEC and crossover regimes in a unified
manner, retrieving the correct limiting behavior
\cite{RanderiaPRB55,RanderiaPRL71}. In this paper we have extended the
path-integral treatment to investigate the vortex state of a fermionic
superfluid, by introducing an appropriate saddle-point for the pairing field.
To set up and solve the resulting saddle point equations, the assumption was
made that the pairing field varies slowly on the scale of the relevant fermion
frequencies. Limitations of this assumption become apparent as one tries to
calculate the exact density profile across the vortex core. Nevertheless, with
this assumption the vortex core size can be calculated as a function of
$1/(k_{F}a_{s})$. The vortex core size varies in the BEC limit according to
the molecular BEC healing length, supporting the formalism. The path-integral
formalism furthermore predicts that the vortex core size (in units of
$1/k_{F}$) tends to a constant value as $1/(k_{F}a_{s})$ is tuned into the BCS regime.

\begin{acknowledgments}
Insightful discussions with A. Pelster are gratefully acknowledged. The
authors also acknowledge discussions with F. Brosens en P. Navez. The Fonds
voor Wetenschappelijk Onderzoek - Vlaanderen provides financial support for
J.T. (`postdoctoraal mandaat') and M.W. (`mandaat aspirant') . This research
has been supported financially by the FWO-V projects Nos. G.0435.03,
G.0306.00, the W.O.G. project WO.025.99N, the GOA BOF UA 2000 UA. J.T.
gratefully acknowledges support of the Special Research Fund of the University
of Antwerp, BOF\ NOI UA 2004.
\end{acknowledgments}

\end{document}